\documentclass{PoS}%
\usepackage{amsmath}
\usepackage{amsfonts}
\usepackage{amssymb}
\usepackage{graphicx}
\usepackage{bm}%
\providecommand{\U}[1]{\protect\rule{.1in}{.1in}}
\title{%
{\vspace{-2cm}%
\begin{flushright}
\small
CHIBA-EP-201\\
KEK Preprint 2012-46
\end{flushright}
\vspace{0.5cm}}
Non-Abelian dual superconductivity and Gluon propagators in the deep IR region for SU(3) Yang-Mills theory %
}

\ShortTitle{non-Abelian dual superconductivity for SU(3) Yang-Mills theory}

\author{\speaker{Akihiro Shibata}\\
        Computing Research Center, High Energy Accelerator Research Organization (KEK), \\
        and Graduate Univ. for Advanced Studies (Sokendai), Tsukuba 305-0801, Japan\\
        E-mail: \email{akihiro.shibata@kek.jp}}
\author{Kei-Ichi Kondo\\
           Department of Physics, Graduate School of Science, Chiba University, Chiba 263-8522, Japan \\
           E-mail: \email{kondok@faculty.chiba-u.jp}}

\author{Seikou Kato\\
           Fukui National College of Technology, Sabae 916-8507, Japan \\
           E-mail: \email{skato@fukui-nct.ac.jp}}
\author{Toru Shinohara\\
           Department of Physics, Graduate School of Science, Chiba University, Chiba 263-8522, Japan \\
           E-mail: \email{sinohara@graduate.chiba-u.jp}}
\abstract{We have proposed the non-Abelian dual superconductivity picture for quark confinement in the SU(3) Yang-Mills (YM) theory, 
and have given numerical evidences for the restricted-field dominance and the non-Abelian magnetic monopole dominance in the string tension 
by applying a new formulation of the YM theory on a lattice. To establish the non-Abelian dual superconductivity picture for quark confinement, 
we have observed the non-Abelian dual Meissner effect in the SU(3) Yang-Mills theory by measuring the chromoelectric flux created 
by the quark-antiquark source, and the non-Abelian magnetic monopole currents induced around the flux. We conclude that 
the dual superconductivity of the SU(3) Yang-Mills theory is strictly the type I and that this type of dual superconductivity is reproduced 
by the restricted field and the non-Abelian magnetic monopole part, in sharp contrast to the SU(2) case: the border of type I and type II.}

\FullConference{Xth Quark Confinement and the Hadron Spectrum,\\
		October 8-12, 2012\\
		TUM Campus Garching, Munich, Germany}

\begin{document}
\section{Introduction}

Quark confinement follows from the area law of the Wilson loop average. The
dual superconductivity is the promising mechanism for quark confinement
\cite{dualSC}. In many preceding works, the Abelian projection \cite{tHooft81}
was used to perform numerical analyses, which exhibited the remarkable results
such as Abelian dominance \cite{Suzuki90}, magnetic monopole dominance
\cite{Stack94-Shiba}, and center vortex dominance \cite{greensite} in the
string tension. However, these results are obtained only in special gauges:
the maximal Abelian (MA) gauge and the Laplacian Abelian gauge within the
Abelian projection, which breaks the gauge symmetry as well as color symmetry
(global symmetry).

In order to overcome the shortcomings of the Abelian projection, we have
presented a new lattice formulation of $SU(N)$ Yang-Mills (YM) theory in the
previous papers \cite{SCGTKKS08L,exactdecomp} (as a lattice version of the
continuum formulations \cite{CFNS-C,KSM05} for $SU(2)$ and \cite{SCGTKKS08}
for $SU(N)$), which gives a decomposition of the gauge link variable suited
for extracting the dominant modes for quark confinement in the gauge
independent way. In the case of $SU(2)$, the decomposition of the gauge link
variable was given on a lattice \cite{NLCVsu2} as a lattice version of the
Cho-Duan-Ge-Faddeev-Niemi decomposition \cite{CFNS-C}. For the gauge group
$G=SU(N)$ ($N\geq3$), it was found that the extension of the decomposition
from $SU(2)$ to $SU(N)$ ($N\geq3$) is not unique and that there are a number
of possible ways of decompositions discriminated by the stability subgroup
$\tilde{H}$ of $G,$ while there is the unique option of $\tilde{H}=U(1)$ in
the $SU(2)$ case \cite{kondo:taira:2000}.

For the case of $G=SU(3)$, in particular, there are two possibilities which we
call the maximal option and the minimal option. The maximal option is obtained
for the stability group $\tilde{H}=U(1)\times U(1)$, which enables us to give
a gauge invariant version of the MA gauge as the Abelian projection
\cite{lattce2007,Suganuma}. The minimal one is obtained for the stability
group $\tilde{H}=U(2)\cong SU(2)\times U(1)$, which is suited for representing
the Wilson loop in the fundamental representation as derived from the
non-Abelian Stokes theorem \cite{KondoNAST}. In the static potential for a
pair of quark and antiquark in the fundamental representation, we have
demonstrated in \cite{NLCVsu3} and \cite{abeliandomSU(3)}: (i) the
restricted-field dominance or \textquotedblleft Abelian\textquotedblright%
\ dominance (which is a gauge-independent (invariant) extension of the
conventionally called Abelian dominance): the string tension $\sigma
_{\mathrm{V}}$ obtained from the decomposed $V$-field (i.e., restricted field)
reproduced the string tension $\sigma_{\mathrm{full}}$ of the original YM
field, $\sigma_{\mathrm{V}}/\sigma_{\mathrm{full}}=93\pm16\%$, (ii) the
gauge-independent non-Abelian magnetic monopole dominance: the string tension
$\sigma_{\mathrm{V}}$ extracted from the restricted field was reproduced by
only the (non-Abelian) magnetic monopole part $\sigma_{\mathrm{mon}}$,
$\sigma_{\mathrm{mon}}/\sigma_{\mathrm{V}}=94\pm9\%$.

In this paper, we give further evidences for establishing the non-Abelian dual
superconductivity picture for quark confinement in $SU(3)$ Yang-Mills theory
claimed in \cite{abeliandomSU(3)} by applying our new formulation to the
$SU(3)$ YM theory on a lattice. First, we study the dual Meissner effect by
measuring the distribution of chromo-flux created by a pair of static quark
and antiquark. We compare the chromo-flux of the original Yang-Mills field
with that of the restricted field and examine if the restricted field
corresponding to the stability group $\tilde{H}=U(2)$ reproduces the dual
Meissner effect, namely, the dominant part of the chromoelectric field
strength of $SU(3)$ Yang-Mills theory. Second, we measure the possible
magnetic monopole current induced around the flux connecting a pair of static
quark and antiquark. Third, we focus on the type of dual superconductivity,
i.e., type I or type II. In the $SU(2)$ case, the extracted field
corresponding to the stability group $\tilde{H}=U(1)$ reproduces the dual
Meissner effect, which gives a gauge invariant version of MA gauge in the
Abelian projection, as will be given in \cite{DualSC:KKSS2012}. In this paper,
we find that the dual superconductivity of the $SU(3)$ Yang-Mills theory is
indeed the type I, in sharp contrast to the $SU(2)$ case: the border of type I
and type II \cite{DualSC:KKSS2012}.

\section{Lattice formulation}

We focus our studies on confinement of quarks in a specific representation,
i.e., the fundamental representation. For this purpose, we consider the Wilson
loop operator for obtaining the quark potential, magnetic monopole current and
chromo-field strength in a gauge invariant way. The Wilson loop operator is
uniquely defined by giving a representation, to which the source quark
belongs. A remarkable fact is that the Wilson loop operator in the fundamental
representation leads us to the minimal option in the sense that it is exactly
rewritten in terms of some of the variables (i.e., the color field
$\mathbf{{n}}$ and the $V$ field) which are the same as those adopted in the
minimal option, as shown in the process of deriving a non-Abelian Stokes
theorem for the Wilson loop operator by Kondo \cite{KondoNAST}. Therefore, we
use the reformulation of the Yang-Mills theory in the minimal option to
calculate the average of the Wilson loop operator in the fundamental
representation. We give a brief summary of a new formulation of the lattice
$SU(3)$ YM theory \cite{SCGTKKS08L,exactdecomp}.

For the original $SU(3)$ gauge link variable $U_{x,\mu}\in SU(3)$, we wish to
decompose it into new variables $V_{x,\mu}$ and $X_{x,\mu}$ which have values
in the $SU(3)$ group, i.e., $X_{x,\mu}\in SU(3)$, $V_{x,\mu}\in SU(3)$,
$U_{x,\mu}=X_{x,\mu}V_{x,\mu}\in SU(3)$, so that $V_{x.\mu}$ could be the
dominant mode for quark confinement, while $X_{x,\mu}$ is the remainder. In
this decomposition, we require that $V_{x,\mu}$ is transformed in the same way
as the original gauge link variable $U_{x,\mu}$ and $X_{x,\mu}$ as a site
variable by the full $SU(3)$ gauge transformation $\Omega_{x}$: $U_{x,\mu
}\longrightarrow U_{x,\nu}^{\prime}=\Omega_{x}U_{x,\mu}\Omega_{x+\mu}^{\dag
},$
\begin{equation}
V_{x,\mu}\longrightarrow V_{x,\nu}^{\prime}=\Omega_{x}V_{x,\mu}\Omega_{x+\mu
}^{\dag},\text{ \ }X_{x,\mu}\longrightarrow X_{x,\nu}^{\prime}=\Omega
_{x}X_{x,\mu}\Omega_{x}^{\dag}.
\end{equation}
We introduce the key variable $\mathbf{h}_{x}$ called the color field. In the
minimal option, the color field is defined by $\mathbf{h}_{x}=\xi(\lambda
^{8}/2)\xi^{\dag}$ $\in Lie[SU(3)/U(2)]$, with $\lambda^{8}$ being the
Gell-Mann matrix and $\xi$ the $SU(3)$ group element. Then, the decomposition
is uniquely determined from Eqs.(\ref{eq:decomp}), if the color field
$\mathbf{h}_{x}$ is specified \cite{exactdecomp}:
\begin{subequations}
\label{eq:decomp}%
\begin{align}
X_{x,\mu}  &  =\widehat{L}_{x,\mu}^{\dag}\det(\widehat{L}_{x,\mu}%
)^{1/3},\text{ \ \ \ }V_{x,\mu}=X_{x,\mu}^{\dag}U_{x,\mu}=g_{x}\widehat
{L}_{x,\mu}U_{x,\mu},\\
\widehat{L}_{x,\mu}  &  =\left(  L_{x,\mu}L_{x,\mu}^{\dag}\right)
^{-1/2}L_{x,\mu},\qquad L_{x,\mu}=\frac{5}{3}\mathbf{1}+\sqrt{\frac{4}{3}%
}(\mathbf{h}_{x}+U_{x,\mu}\mathbf{h}_{x+\mu}U_{x,\mu}^{\dag})+8\mathbf{h}%
_{x}U_{x,\mu}\mathbf{h}_{x+\mu}U_{x,\mu}^{\dag}\text{ .}%
\end{align}
In order to determine the configuration $\{\mathbf{h}_{x}\}$ of color fields,
we use the reduction condition \cite{SCGTKKS08L,exactdecomp} which guarantees
that the new theory written in terms of new variables ($X_{x,\mu}$,$V_{x,\mu}%
$) is equipollent to the original YM theory. Here, we use the reduction
condition: for a given configuration of the original link variables $U_{x,\mu
}$, color fields $\left\{  \mathbf{h}_{x}\right\}  $ are obtained by
minimizing the functional:
\end{subequations}
\begin{equation}
F_{\text{red}}[\{\mathbf{h}_{x}\}]=\sum_{x,\mu}\mathrm{tr}\left\{  (D_{\mu
}^{\epsilon}[U_{x,\mu}]\mathbf{h}_{x})^{\dag}(D_{\mu}^{\epsilon}[U_{x,\mu
}]\mathbf{h}_{x})\right\}  . \label{eq:reduction}%
\end{equation}

\section{Method and results}

We generate configurations of the YM gauge link variable $\{U_{x,\mu}\}$ using
the standard Wilson action on a $24^{4}$ lattice at $\beta=6.2$. The gauge
link decomposition is obtained according to the framework given in the
previous section: the color field configuration $\{h_{x}\}$ is obtained by
solving the reduction condition of minimizing the functional
eq.(\ref{eq:reduction}) for each gauge configuration $\{U_{x,\mu}\}$, and then
the decomposed variables $\{V_{x,\mu}\}$, $\{X_{x,\mu}\}$ are obtained by
using the formula eq.(\ref{eq:decomp}). In the measurement of the Wilson loop
average, we apply the APE smearing technique to reduce noises \cite{APEsmear}.

\subsection{Dual Meissner effect}

\begin{figure}[ptb]
\begin{center}
\vspace{-6mm}\ \ \ \ \begin{minipage}{3.8cm}
\includegraphics[
height=3.7cm
]
{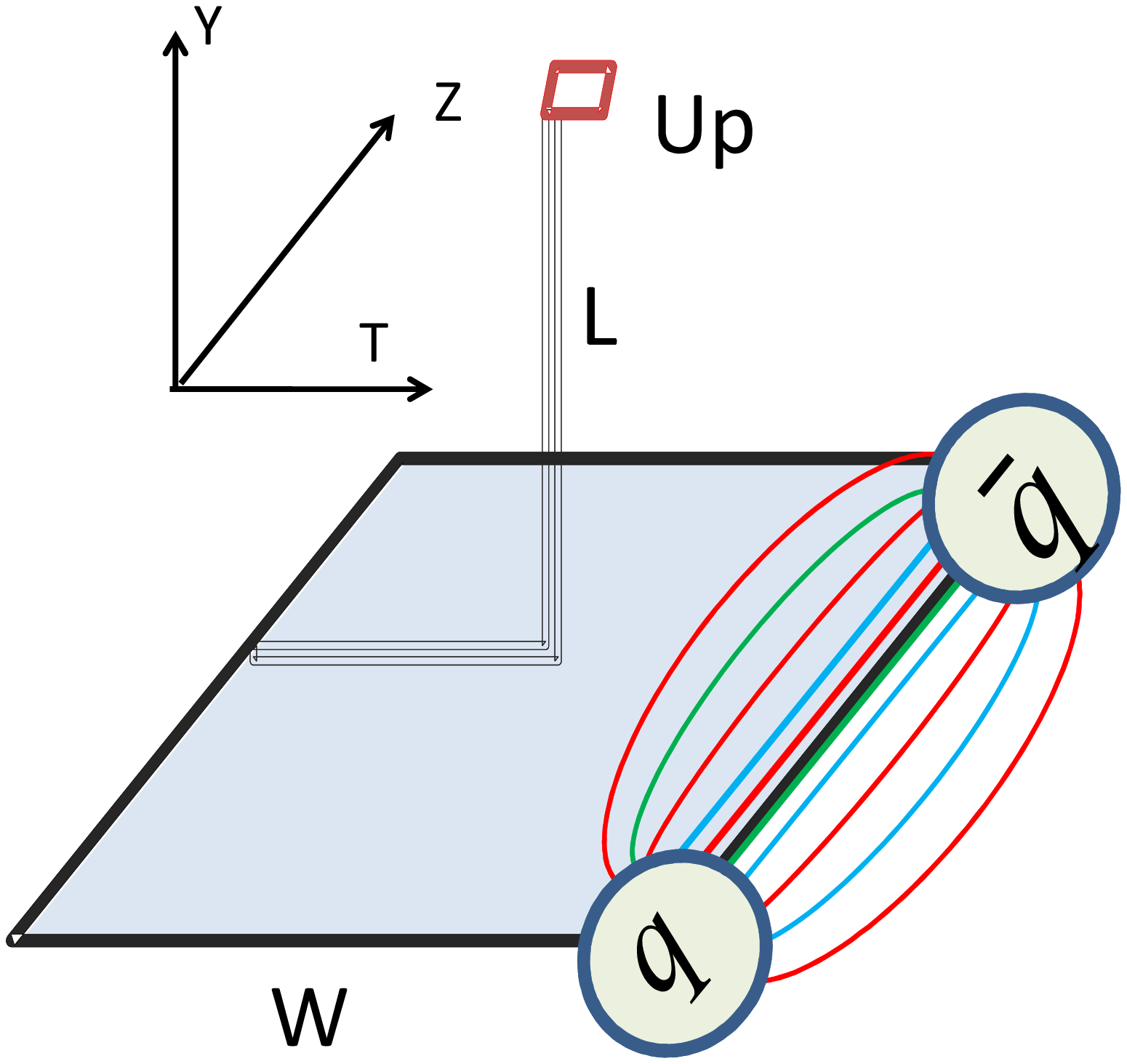}
\end{minipage}\includegraphics[
width=4.4cm,
angle=270,
origin=b
]
{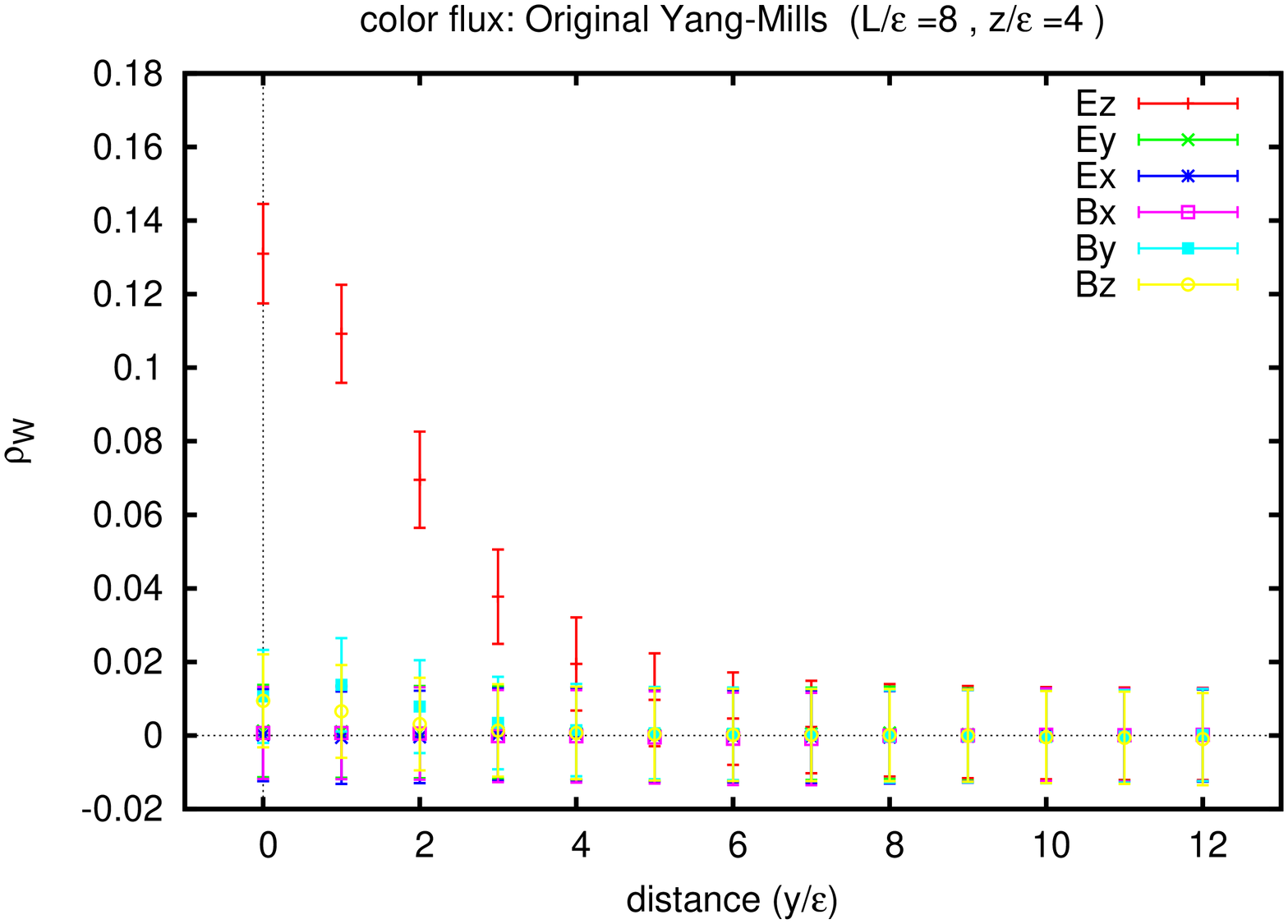} \includegraphics[
width=4.4cm,
angle=270,
origin=b
]
{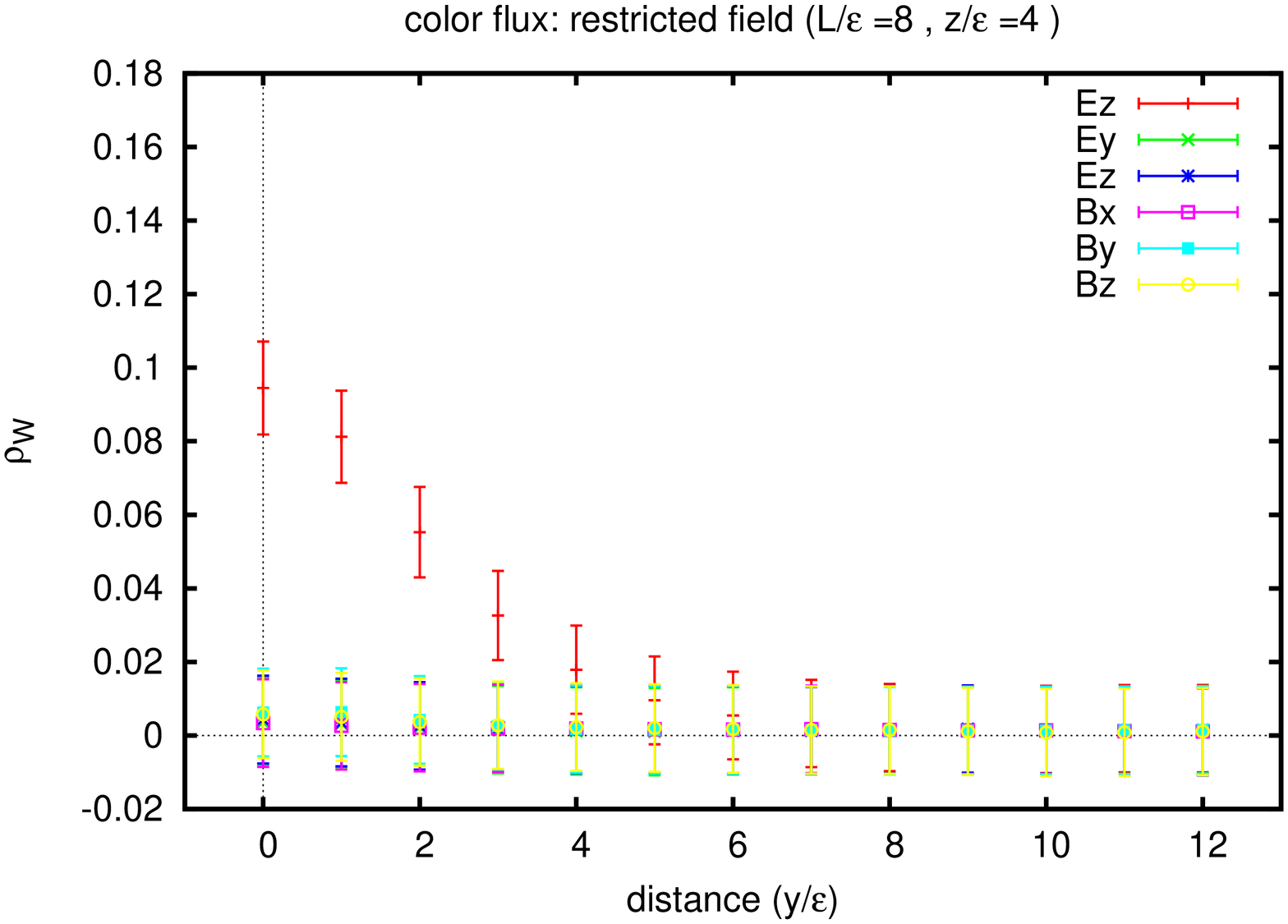}
\end{center}
\caption{ Measurement of components of the chromoelectric field $\bm{E}$ and
chromomagnetic field $\bm{B}$ as functions of the distance $y$ from the $z$
axis. (left) The gauge invariant connected correlator ($U_{p}LWL^{\dag}$)
between plaquette U \ and Wilson loop W. (Center panel) the original $SU(3)$
YM field, (Right panel) the restricted  field. }%
\label{fig:measure}%
\end{figure}

We investigate the non-Abelian dual Meissner effect as the mechanism of quark
confinement. In order to extract the chromo-field, we use a gauge-invariant
correlation function proposed in \cite{DiGiacomo:1990hc}: The chromo-field
created by a quark-antiquark pair in $SU(N)$ Yang-Mills theory is measured by
using a gauge-invariant connected correlator between a plaquette and the
Wilson loop (see left panel of Fig.\ref{fig:measure}):%
\begin{equation}
F_{\mu\nu}[U]:=\epsilon^{-2}\sqrt{\frac{\beta}{2N}}\rho_{W},\text{\qquad}%
\rho_{W}:=\frac{\left\langle \mathrm{tr}\left(  U_{P}L^{\dag}WL\right)
\right\rangle }{\left\langle \mathrm{tr}\left(  W\right)  \right\rangle
}-\frac{1}{N}\frac{\left\langle \mathrm{tr}\left(  U_{P}\right)
\mathrm{tr}\left(  W\right)  \right\rangle }{\left\langle \mathrm{tr}\left(
W\right)  \right\rangle }, \label{eq:Op}%
\end{equation}
where $F_{\mu\nu}[U]$ is the gauge-invariant chromo-field strength,
$\beta:=2N/g^{2}$ the lattice gauge coupling constant, $W$ the Wilson loop in
$Z$-$T$ plane representing a pair of quark and antiquark, $U_{P}$ a plaquette
variable as the probe operator to measure the chromo-field strength at the
point $P$, and $L$ the Wilson line connecting the source $W$ and the probe
$U_{P}$. Here $L$ is necessary to guarantee the gauge invariance of the
correlator $\rho_{W}$ and hence the probe is identified with $LU_{P}%
L^{\dagger}$.
The symbol $\left\langle \mathcal{O}\right\rangle $ denotes the average of the
operator $\mathcal{O}$ in the space and the ensemble of the configurations.

We measure correlators between the plaquette $U_{P}$ and the chromo-field
strength of the restricted field $V_{x,\mu}$ as well as the original YM field
$U_{x,\mu}$. See the center panel of Fig.~\ref{fig:measure}. Here the quark
and antiquark source is introduced as $8\times8$ Wilson loop ($W$) in the
$Z$-$T$ plane, and the probe $(U_{p})$ is set at the center of the Wilson loop
and moved along the $Y$-direction. The center and right panel of
Fig.~\ref{fig:measure} show respectively the results of measurements for the
chromoelectric and chromomagnetic fields $F_{\mu\nu}[U]$ for the original
$SU(3)$ field $U$ and $F_{\mu\nu}[V]$ for the restricted field $V$, where the
field strength $F_{\mu\nu}[V]$ is obtained by using $V_{\,x,\mu}$ in
eq(\ref{eq:Op}) instead of $U_{x,\mu}$. We have checked that even if $W[U]$ is
replaced by $W[V]$, together with replacement of the probe $LU_{P}L^{\dagger}$
by the corresponding $V$ version, the change in the magnitude of the field
strength $F_{\mu\nu}$ remains within at most a few percent.

From Fig.\ref{fig:measure} we find that only the $E_{z}$ component of the
chromoelectric field $(E_{x},E_{y},E_{z})=(F_{10},F_{20},F_{30})$ connecting
$q$ and $\bar{q}$ has non-zero value for both the restricted field $V$ and the
original YM field $U$. The other components are zero consistently within the
numerical errors. This means that the chromomagnetic field $(B_{x},B_{y}%
,B_{z})=(F_{23},F_{31},F_{12})$ connecting $q$ and $\bar{q}$ does not exist
and that the chromoelectric field is parallel to the $z$ axis on which quark
and antiquark are located. The magnitude $E_{z}$ quickly decreases in the
distance $y$ away from the Wilson loop.\begin{figure}[ptb]
\begin{center}
\vspace{-6mm}\includegraphics[
height=68mm,
angle=270
]
{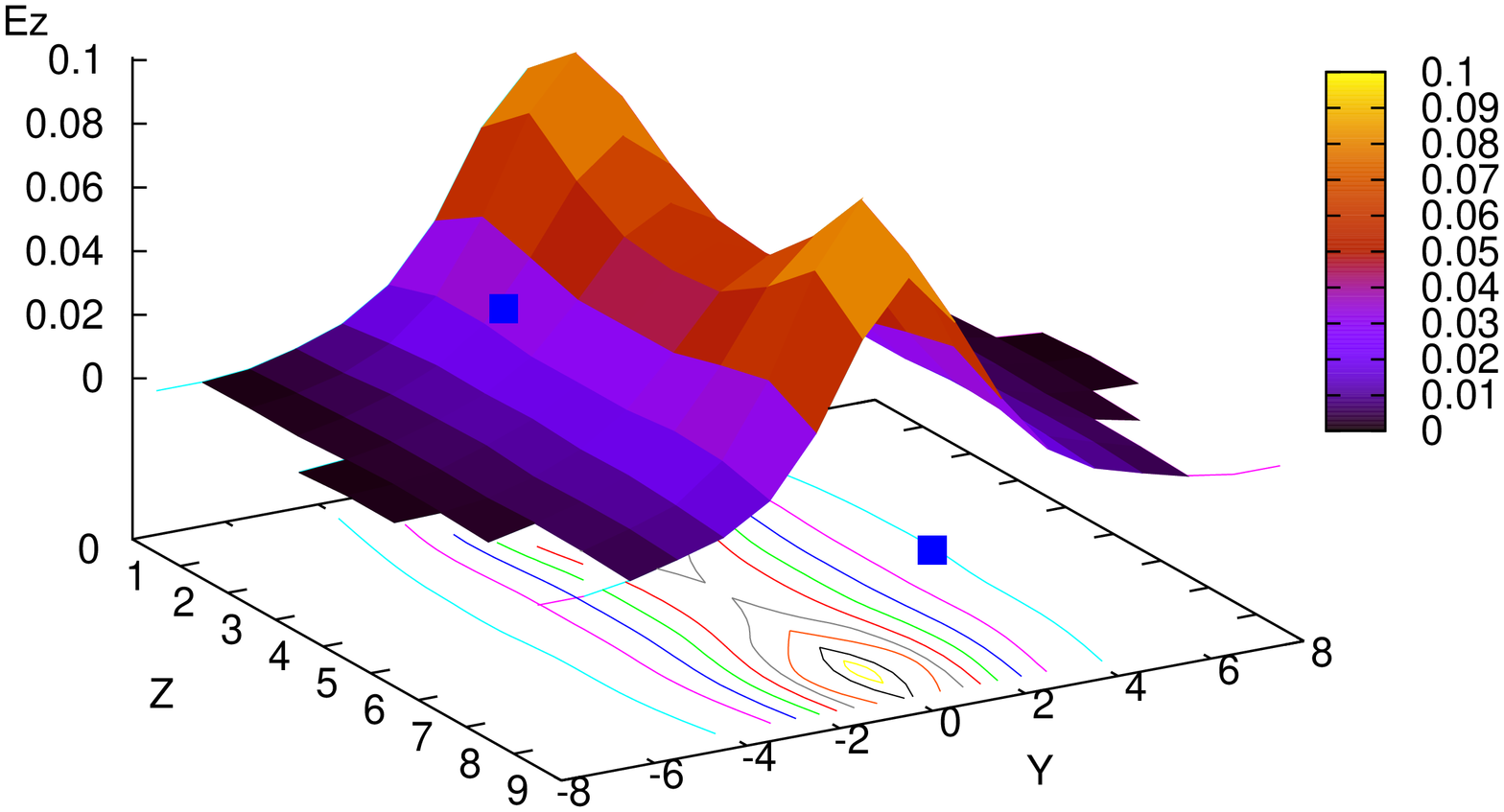} \ \ \ \ \quad\includegraphics[
height=68mm,
angle=270
]
{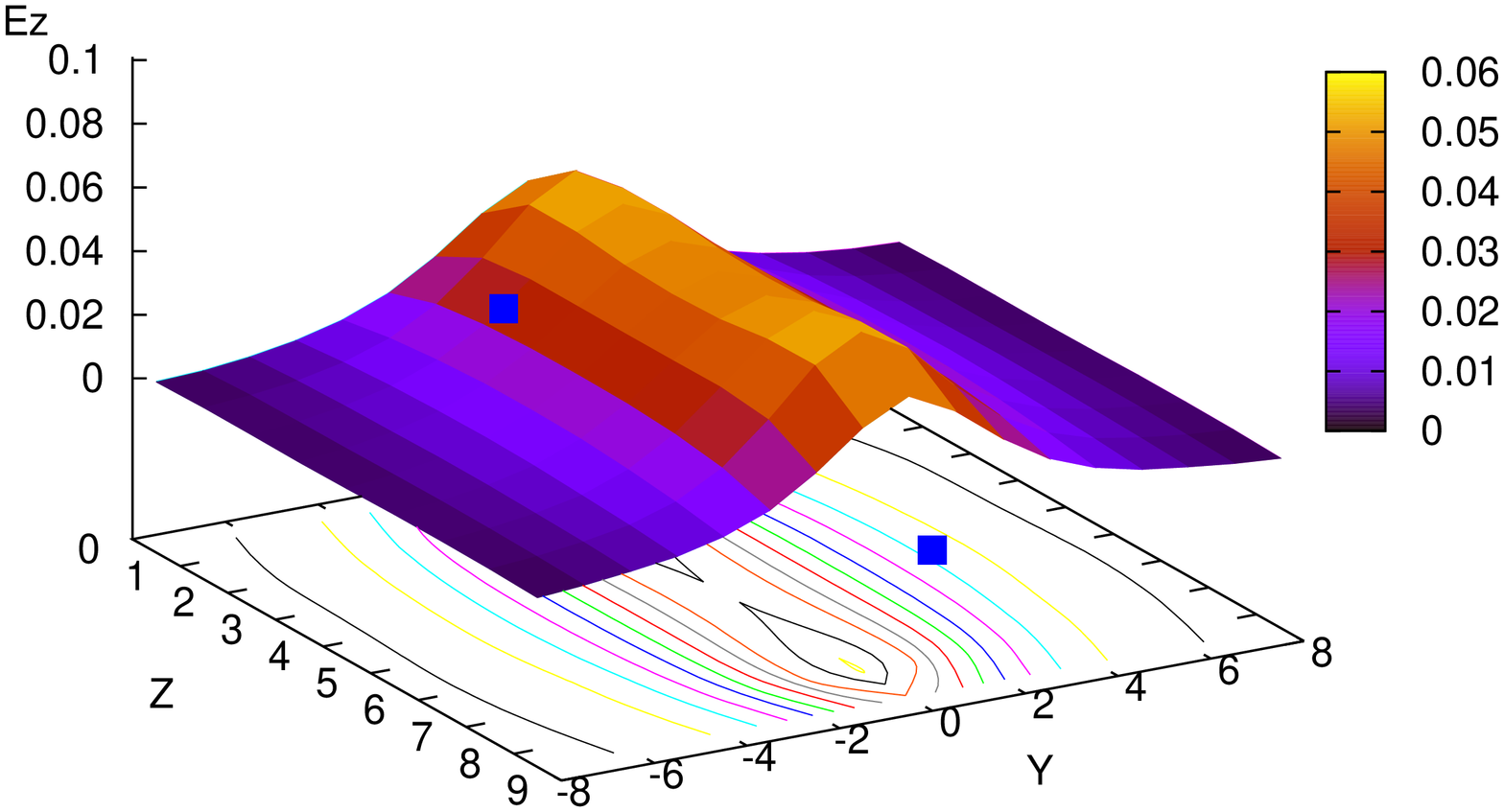} \vspace{-5mm}
\end{center}
\caption{ The distribution in $Y$-$Z$ plane of the chromoelectric field
$E_{z}$ connecting a pair of quark and antiquark: (Left panel) chromoelectric
field produced from the original YM field, (Right panel) chromoelectric field
produced from the restricted field. }%
\label{fig:fluxtube}%
\end{figure}

To see the profile of the nonvanishing component $E_{z}$ of the chromoelectric
field in detail, we explore the distribution of chromoelectric field on the
2-dimensional plane. Fig.~\ref{fig:fluxtube} shows the distribution of $E_{z}$
component of the chromoelectric field, where the quark-antiquark source
represented as $9\times11$ Wilson loop $W$ is placed at $(Y,Z)=(0,0),(0,9)$,
and the probe $U$ is displaced on the $Y$-$Z$ plane at the midpoint of the
$T$-direction. The position of a quark and an antiquark is marked by the solid
(blue) box. The magnitude of $E_{z}$ is shown by the height of the 3D plot and
also the contour plot in the bottom plane. The left panel of
Fig.~\ref{fig:fluxtube} shows the plot of $E_{z}$ for the $SU(3)$ YM field
$U$, and the right panel of Fig.~\ref{fig:fluxtube} for the restricted-field
$V$. We find that the magnitude $E_{z}$ is quite uniform for the restricted
part $V$, while it is almost uniform for the original part $U$ except for the
neighborhoods of the locations of $q$, $\bar{q}$ source. This difference is
due to the contributions from the remaining part $X$ which affects only the
short distance, as will be discussed in the next section.

\subsection{Magnetic current}

Next, we investigate the relation between the chromoelectric flux and the
magnetic current.
The magnetic(-monopole) current can be calculated as
\begin{equation}
\mathbf{k}={}^{\ast}dF[\mathbf{V}] , \label{def-k}%
\end{equation}
where $F[\mathbf{V}]$ is the field strength (2-form) of the restricted field
(1-form) $\mathbf{V}$, $d$ the exterior derivative and $^{\ast}$ denotes the
Hodge dual operation. Note that non-zero magnetic current follows from
violation of the Bianchi identity (If the field strength was given by the
exterior derivative of $\mathbf{V}$ field (one-form), $F[\mathbf{V}%
]=d\mathbf{V}$, \ we would obtain $\mathbf{k=}^{\ast}d^{2}\mathbf{V}$ $=0$).

\begin{figure}[ptb]
\begin{center}
\vspace{-2mm} \includegraphics[
width=4.5cm
]
{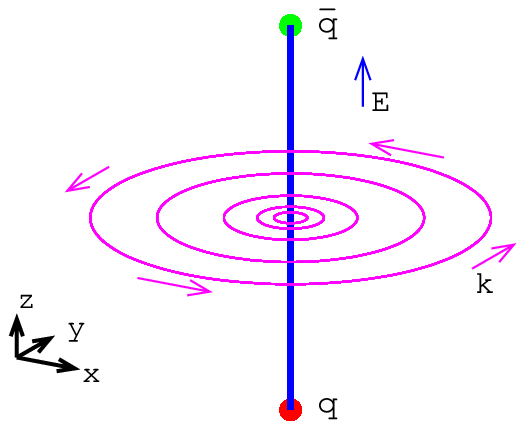} \quad\ \includegraphics[
width=66mm
]
{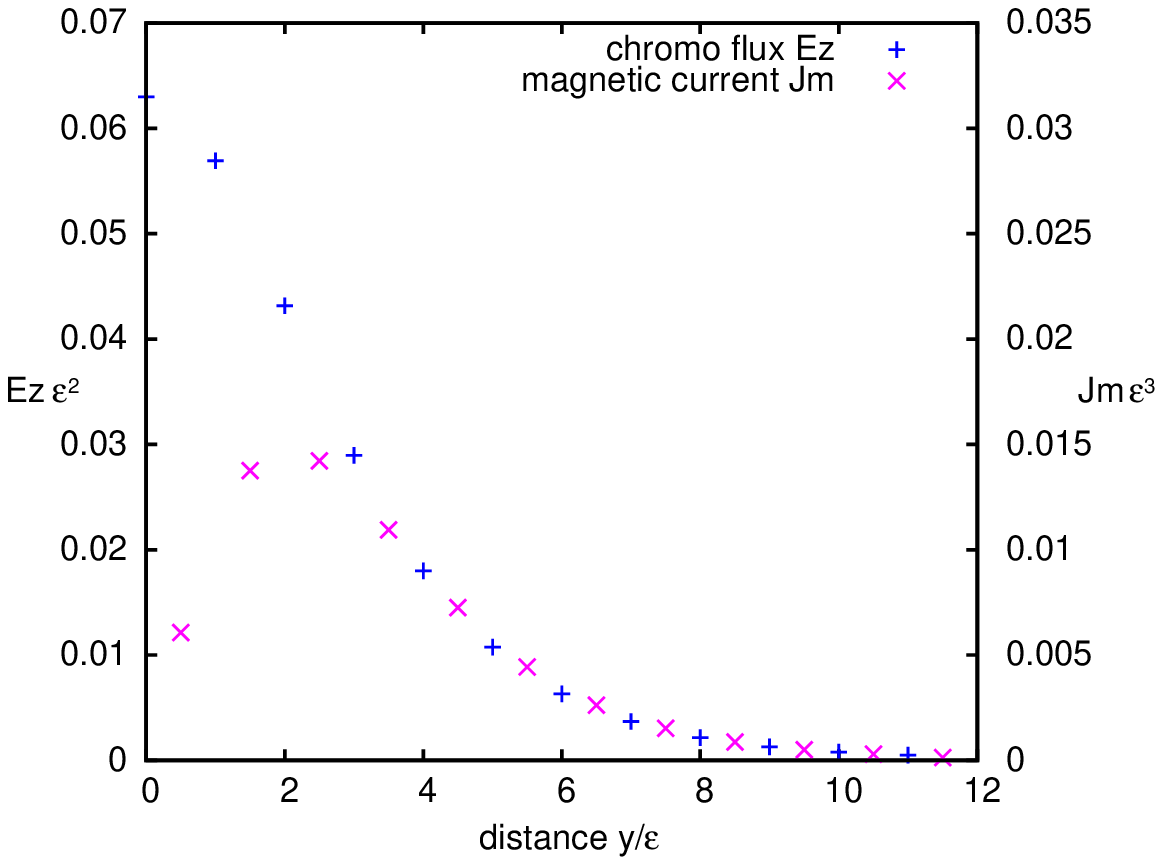}
\end{center}
\par
\vspace{-5mm} \caption{{}The magnetic-monopole current $\mathbf{k}$ induced
around the flux along the $z$ axis connecting a quark-antiquark pair. (Left
panel) The positional relationship between the chromoelectric field $E_{z}$
and the magnetic current $\mathbf{k}$. (Right panel) The magnitude of the
chromo-electronic current $E_{z}$ and the magnetic current $J_{m}%
=|\mathbf{k}|$ as functions of the distance $y$ from the $z$ axis. }%
\label{fig:Mcurrent}%
\end{figure}

Fig.~\ref{fig:Mcurrent} shows the magnetic current measured in $X$-$Y$ plane
at the midpoint of quark and antiquark pair in the $Z$-direction. The left
panel of Fig.~\ref{fig:Mcurrent} shows the positional relationship between
chromoelectric flux and magnetic current. The right panel of
Fig.~\ref{fig:Mcurrent} shows the magnitude of the chromoelectric field
$E_{z}$ (left scale) and the magnetic current $k$ (right scale). The existence
of nonvanishing magnetic current $k$ around the chromoelectric field $E_{z}$
supports the dual picture of the ordinary superconductor exhibiting the
electric current $J$ around the magnetic field $B$.

In our formulation, it is possible to define a gauge-invariant
magnetic-monopole current $k_{\mu}$ by using $V$-field,
which is obtained from the field strength $\mathcal{F}_{\mu\nu}[\mathbf{V}]$
of the field $\mathbf{V}$, as suggested from the non-Abelian Stokes theorem
\cite{KondoNAST}. It should be also noticed that this magnetic-monopole
current is a non-Abelian magnetic monopole extracted from the $V$ field, which
corresponds to the stability group $\tilde{H}=U(2)$. The magnetic-monopole
current $k_{\mu}$ defined in this way can be used to study the magnetic
current around the chromoelectric flux tube, instead of the above definition
of $k$ (\ref{def-k}). The comparison of two monopole currents $k$ will be done
in the forthcoming paper.

\subsection{Type of dual superconductivity}

\begin{figure}[ptb]
\begin{center}
\vspace{-6mm} \includegraphics[
width=40mm,
angle=270
]
{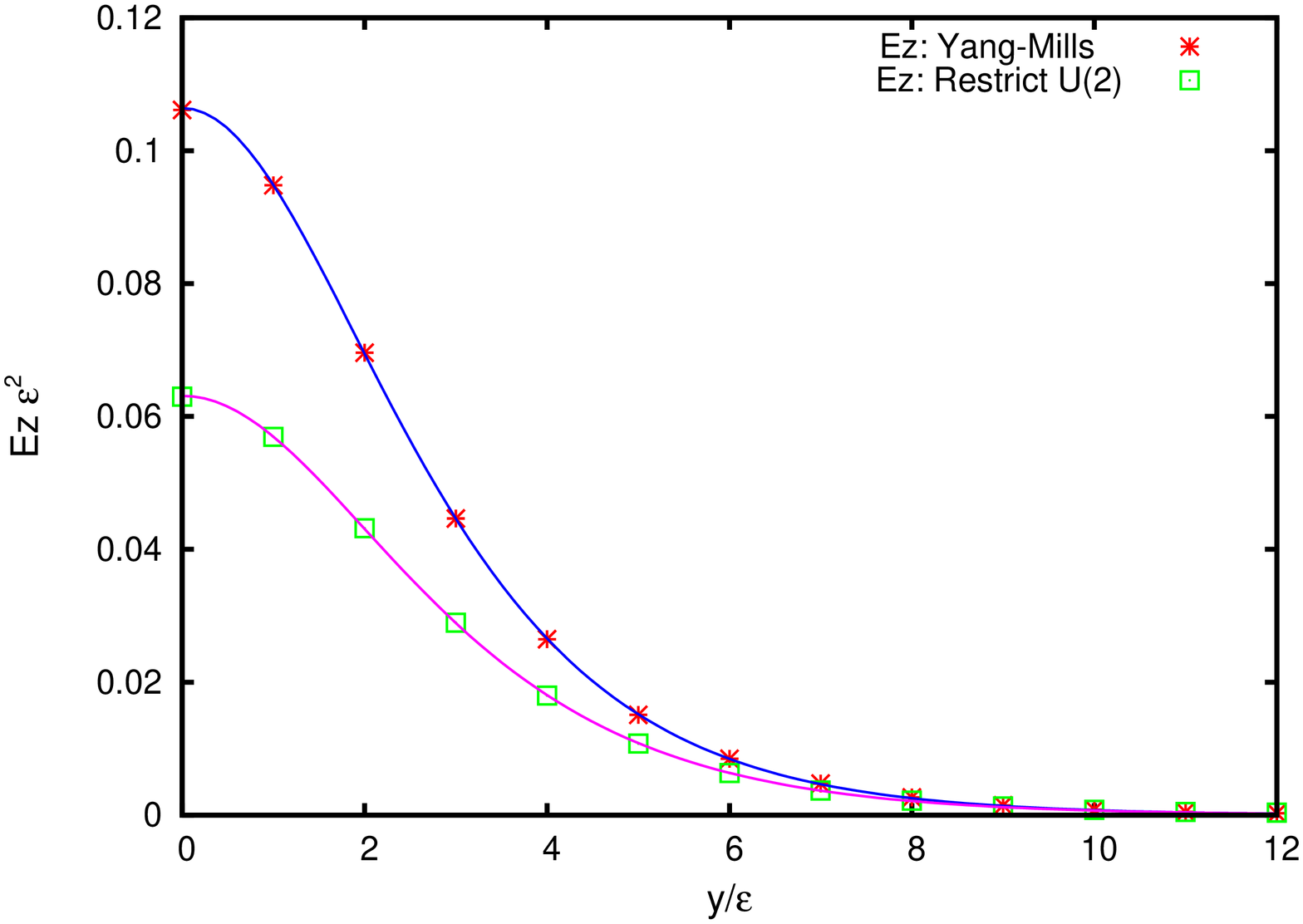} \qquad\ \includegraphics[
width=43mm,
angle=270
]
{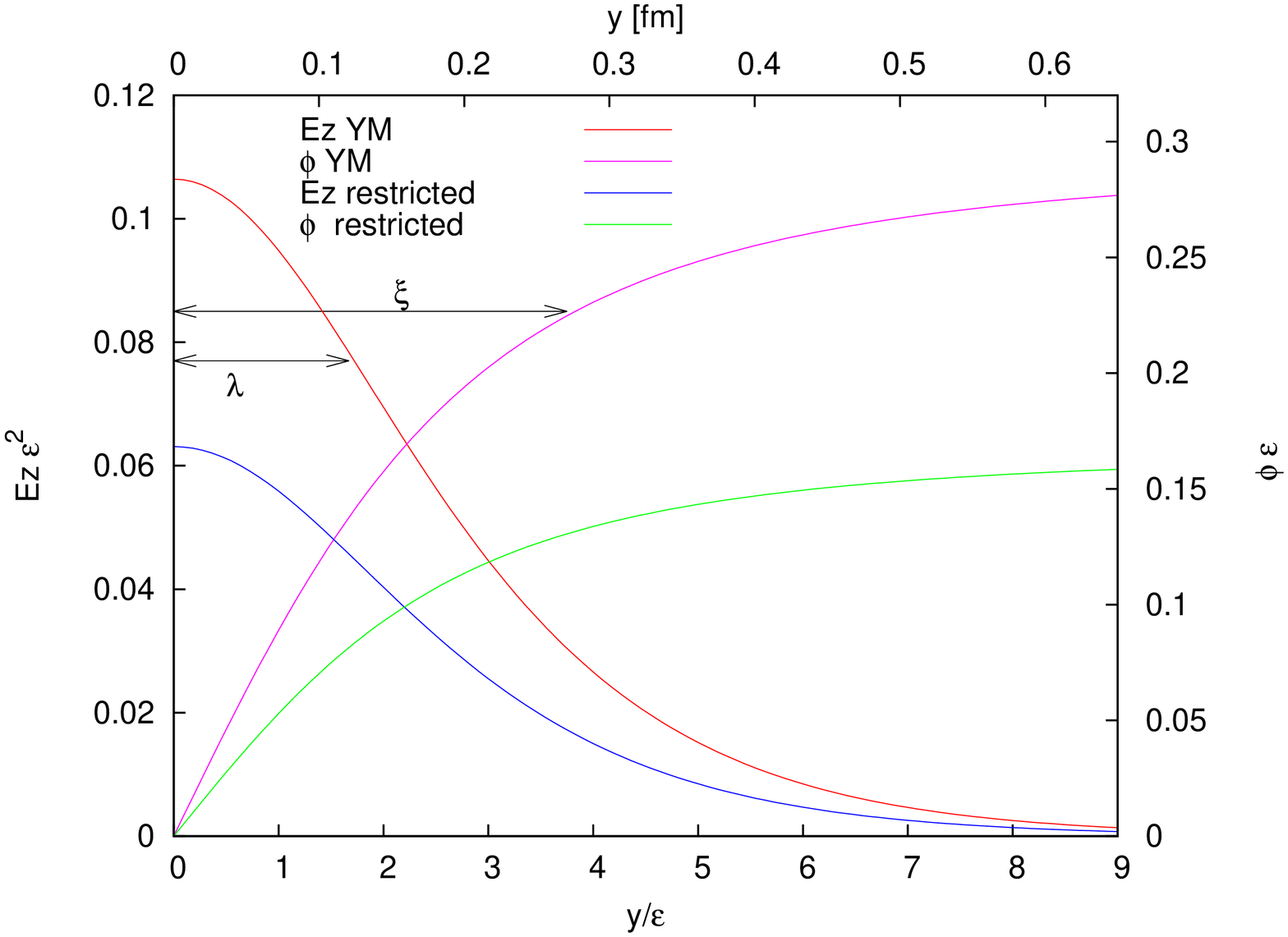} \vspace{-5mm}
\end{center}
\caption{ (Left panel) The plot of the chromoelectric field $E_{z}$ versus the
distance $y$ in units of the lattice spacing $\epsilon$ and the fitting as a
function $E_{z}(y)$ of $y$ according to ( {\protect\ref{fitting}}). The red cross for
the original $SU(3)$ field and the green square symbol for the restricted
field. (Right panel) The order parameter $\phi$ reproduced as a function
$\phi(y)$ of $y$ according to ({\protect\ref{order-f}}), togather with the
chromoelectric field $E_{z}(y)$. }%
\label{fig:type}%
\end{figure}

Moreover, we investigate the QCD vacuum, i.e., type of the dual
superconductor. The left panel of Fig.\ref{fig:type} is the plot for the
chromoelectric field $E_{z}$ as a function of the distance $y$ in units of the
lattice spacing $\epsilon$ for the original $SU(3)$ field and for the
restricted field.

In order to examine the type of the dual superconductivity, we apply the
formula for the magnetic field derived by Clem \cite{Clem75} in the ordinary
superconductor based on the Ginzburg-Landau (GL) theory to the chromoelectric
field in the dual superconductor. In the GL theory, the gauge field $A$ and
the scalar field $\phi$ obey simultaneously the GL equation and the Ampere
equation:
\begin{align}
(\partial^{\mu}-iqA^{\mu})(\partial_{\mu}-iqA_{\mu})\phi+\lambda(\phi^{\ast
}\phi-\eta^{2})  &  =0,\\
\partial^{\nu}F_{\mu\nu}+iq[\phi^{\ast}(\partial_{\mu}\phi-iqA_{\mu}%
\phi)-(\partial_{\mu}\phi-iqA_{\mu}\phi)^{\ast}\phi]  &  =0.
\end{align}

Usually, in the dual superconductor of the type II, it is justified to use the
asymptotic form $K_{0}(y/\lambda)$ to fit the chromoelectric field in the
large $y$ region (as the solution of the Ampere equation in the dual GL
theory). However, it is clear that this solution cannot be applied to the
small $y$ region, as is easily seen from the fact that $K_{0}(y/\lambda)
\to\infty$ as $y \to0$. In order to see the difference between type I and type
II, it is crucial to see the relatively small $y$ region. Therefore, such a
simple form cannot be used to detect the type I dual superconductor. However,
this important aspect was ignored in the preceding studies except for a work
\cite{Cea:2012qw}.

On the other hand, Clem \cite{Clem75} does not obtain the analytical solution
of the GL equation explicitly and use an approximated form for the scalar
field $\phi$ (given below in ({\protect\ref{order-f}})). This form is used to solve the
Ampere equation exactly to obtain the analytical form for the gauge field
$A_{\mu}$ and the resulting magnetic field $B$. This method does not change
the behavior of the gauge field in the long distance, but it gives a finite
value for the gauge field even at the origin. Therefore, we can obtain the
formula which is valid for any distance (core radius) $y$ from the axis
connecting $q$ and $\bar{q}$: the profile of chromoelectric field in the dual
superconductor is obtained:
\begin{equation}
E_{z}(y)=\frac{\Phi}{2\pi}\frac{1}{\zeta\lambda}\frac{K_{0}(R/\lambda)}%
{K_{1}(\zeta/\lambda)},\text{ }R=\sqrt{y^{2}+\zeta^{2}}, \label{eq:fluxClem}%
\end{equation}
provided that the scalar field is given by (See the right panel of
Fig.\ref{fig:type})
\begin{equation}
\phi(y)=\frac{\Phi}{2\pi}\frac{1}{\sqrt{2}\lambda}\frac{y}{\sqrt{y^{2}%
+\zeta^{2}}}, \label{order-f}%
\end{equation}
where $K_{\nu}$ is the modified Bessel function of the $\nu$-th order,
$\lambda$ the parameter corresponding to the London penetration length,
$\zeta$ a variational parameter for the core radius, and $\Phi$ external
electric flux. In the dual superconductor, we define the GL parameter $\kappa$
as the ratio of the London penetration length $\lambda$ and the coherence
length $\xi$ which measures the coherence of the magnetic monopole condensate
(the dual version of the Cooper pair condensate):$\kappa=\lambda/\xi.$ \ It is
given by \cite{Clem75}
\begin{equation}
\kappa=\sqrt{2}\frac{\lambda}{\zeta}\sqrt{1-K_{0}^{2}(\zeta/\lambda)/K_{1}%
^{2}(\zeta/\lambda)}.
\end{equation}

\begin{table}[t]%
\begin{tabular}
[c]{|l||c|c|c||c|c|c|c|c|}\hline
& $a\epsilon^{2}$ & $b\epsilon$ & $c$ & $\lambda/\epsilon$ & $\zeta/\epsilon$
& $\xi/\epsilon$ & $\Phi$ & $\kappa$\\\hline
SU(3) YM field & $0.804(4)$ & $0.598(5)$ & $1.878(4)$ & $1.672(14)$ &
$3.14(9)$ & $3.75(12)$ & $4.36(30)$ & $0.45(1)$\\\hline
restricted field & $0.435(3)$ & $0.547(7)$ & $1.787(5)$ & $1.828(23)$ &
$3.26(13)$ & $3.84(19)$ & $2.96(30)$ & $0.48(2)$\\\hline
\end{tabular}
\caption{ The properties of the Yang-Mills vacuum as the dual superconductor
obtained by fitting the data of chromoelectric field with the prediction of
the dual Ginzburg-Landau theory. }%
\label{Table:GL-fit}%
\end{table}

According to the formula Eq.(\ref{eq:fluxClem}), we estimate the GL parameter
$\kappa$ for the dual superconductor of $SU(3)$ YM theory, although this
formula is obtained for the ordinary superconductor of $U(1)$ gauge field. By
using the fitting function:
\begin{equation}
E(y)=aK_{0}(\sqrt{b^{2}y^{2}+c^{2}}),\quad a=\frac{\Phi}{2\pi}\frac{1}%
{\zeta\lambda}\frac{1}{K_{1}(\zeta/\lambda)},\quad b=\frac{1}{\lambda},\quad
c=\frac{\zeta}{\lambda}, \label{fitting}%
\end{equation}
we obtain the result shown in Table~\ref{Table:GL-fit}. The superconductor is
type I if $\kappa<\kappa_{c}$, while type II if $\kappa>\kappa_{c}$, where the
critical value of GL parameter dividing the type of the superconductor is
given by $\kappa_{c}=1/\sqrt{2}\simeq0.707$. Our data clearly shows that the
dual superconductor of $SU(3)$ Yang-Mills theory is type I with
\begin{equation}
\kappa=0.45\pm0.01\text{ ,\qquad\thinspace}\lambda=0.1207\pm0.017\text{ fm
,\qquad}\xi=0.2707\pm0.086\text{ fm, }%
\end{equation}
where we have used the string tension $\sigma_{\text{phys}}=(440\text{MeV}%
)^{2}$, and data of lattice spacing is taken from the TABLE I in
Ref.\cite{Edward98}. \ This result is consistent with a quite recent result
obtained independently by Cea, Cosmai and Papa \cite{Cea:2012qw}. Moreover,
our result shows that the restricted part plays the dominant role in
determining the type of the non-Abelian dual superconductivity of the $SU(3)$
Yang-Mills theory, i.e., type I with
\begin{equation}
\kappa=0.48\pm0.02\text{ ,}\qquad\lambda=0.132\pm0.03\text{ fm , \qquad}%
\xi=0.277\pm0.014\text{ fm}.
\end{equation}
This is a novel feature overlooked in the preceding studies. Thus the
restricted-field dominance can be seen also in the determination of the type
of dual superconductivity where the discrepancy is just the normalization of
the chromoelectric field at the core $y=0$, coming from the difference of the
total flux $\Phi$. These are gauge-invariant results. Note again that this
restricted-field and the non-Abelian magnetic monopole extracted from it
reproduce the string tension in the static quark--antiquark potential
\cite{abeliandomSU(3)}.

Our result should be compared with the result obtained by using the Abelian
projection: Y.~Matsubara et. al \cite{Matsubara:1993nq} suggests
$\kappa=0.5\sim1$(which is $\beta$ dependent), border of type I and type II
for both $SU(2)$ and $SU(3)$. In $SU(2)$ case, on the other hand, there are
other works \cite{Suzuki04:Chelnodub05} which conclude that the type of vacuum
is at the border of type I and type II. We should mention the work
\cite{Cardoso:2010kw} which concludes that the dual superconductivity of
$SU(3)$ Yang-Mills theory is type II with $\kappa=1.2\sim1.3$. This conclusion
seems to contradict our result for $SU(3)$. If the above formula
(\ref{eq:fluxClem}) is applied to the data of \cite{Cardoso:2010kw}, we have
the same conclusion, namely, the type I with $\kappa=0.47\sim0.50$.
Therefore, the data obtained in \cite{Cardoso:2010kw} are consistent with
ours. The difference between type I and type II is attributed to the way of
fitting the data with the formula for the chromo-field.


\section{Summary and outlook}

We have given further numerical evidences for confirming the non-Abelian dual
superconductivity for $SU(3)$ YM theory proposed in \cite{abeliandomSU(3)}.
For this purpose, we have used our new formulation of $SU(3)$ YM theory on a
lattice \cite{SCGTKKS08L,exactdecomp} to extract the restricted field from the
original $SU(3)$ YM field, which has played a dominant role in confinement of
quarks in the fundamental representation, i.e., the restricted-field dominance
and the non-Abelian magnetic monopole dominance in the string tension, as
shown in the previous studies \cite{abeliandomSU(3)}.

We have focused on the dual Meissner effect and have measured the
chromoelectric field connecting a quark and an antiquark for both the original
YM field and the restricted field. We have observed the dual Meissner effect
in $SU(3)$ YM theory, i.e., only the chromoelectric field exists and the
magnetic-monopole current is induced around the flux connecting a quark and an
antiquark. Moreover, we have determined the type of non-Abelian dual
superconductivity, i.e., type I for $SU(3)$ YM theory, which should be
compared with the border of type I and II for the SU(2) YM theory. These
features are reproduced only from the restricted part. These results confirm
the non-Abelian dual superconductivity picture for quark confinement.


\subsection*{Acknowledgement}

This work is supported by Grant-in-Aid for Scientific Research (C) 24540252
from the Japan Society for the Promotion Science (JSPS), and also in part by
JSPS Grant-in-Aid for Scientific Research (S) 22224003. The numerical
calculations are supported by the Large Scale Simulation Program No.T11-15
(FY2011) and No.12-13 (FY2012) of High Energy Accelerator Research
Organization (KEK).

\end{document}